\documentclass[%
 aip,
 amsmath,amssymb,
 reprint,%
]{revtex4-1}

\usepackage{graphicx}% Include figure files
\usepackage{dcolumn}% Align table columns on decimal point
\usepackage{bm}% bold math
\usepackage[utf8]{inputenc}
\usepackage[T1]{fontenc}
\usepackage{mathptmx}
\usepackage{etoolbox}
\makeatletter
\def\@email#1#2{%
 \endgroup
 \patchcmd{\titleblock@produce}
  {\frontmatter@RRAPformat}
  {\frontmatter@RRAPformat{\produce@RRAP{*#1\href{mailto:#2}{#2}}}\frontmatter@RRAPformat}
  {}{}
}%
\makeatother
\begin{document}
\preprint{AIP/123-QED}

\title[High-resolution nanopatterning of NbN thin films]{High-resolution top-down nanopatterning of NbN thin films for superconducting metamaterials}

\author{Francesca Incalza}
\email{incal723@mit.edu}
\altaffiliation{These authors contributed equally to this work.}
\affiliation{Massachusetts Institute of Technology, Department of Electrical Engineering and Computer Science\\ Cambridge, MA, USA}

\author{Davide Mondin}
\email{mondin@mit.edu}
\altaffiliation{These authors contributed equally to this work.}
\affiliation{Massachusetts Institute of Technology, Department of Electrical Engineering and Computer Science\\ Cambridge, MA, USA}

\author{Dip Joti Paul}
\affiliation{Massachusetts Institute of Technology, Department of Electrical Engineering and Computer Science\\ Cambridge, MA, USA}

\author{Matteo Castellani}
\affiliation{Massachusetts Institute of Technology, Department of Electrical Engineering and Computer Science\\ Cambridge, MA, USA}

\author{Alejandro Simon}
\affiliation{Massachusetts Institute of Technology, Department of Electrical Engineering and Computer Science\\ Cambridge, MA, USA}

\author{Rocco A. Vitalone}
\affiliation{Department of Physics, Columbia University, New York, NY, USA}

\author{Dmitri N. Basov}
\affiliation{Department of Physics, Columbia University, New York, NY, USA}

\author{Karl K. Berggren}
\affiliation{Massachusetts Institute of Technology, Department of Electrical Engineering and Computer Science\\ Cambridge, MA, USA}

\begin{abstract}

Periodic nanopatterning offers a route to tailoring vortex motion and the electromagnetic response of superconducting films, providing a basis for superconducting metamaterials. Here, we report a high-resolution top-down nanopatterning process for ultrathin epitaxial NbN films on sapphire substrates. Electron-beam lithography using hydrogen silsesquioxane (HSQ) and ZEP 530A resists was optimized through proximity-effect correction and adhesion-layer integration. Dense superconducting nanowire grids with pitches down to 70 nm were fabricated, alongside millimeter-square areas patterned with 100-nm-pitch hole arrays with high pattern fidelity. A thin Si capping layer was found to significantly improve HSQ adhesion on epitaxial NbN. The nanopatterns were successfully transferred into $\sim$5-nm-thick NbN films by reactive ion etching (RIE). Structural and dimensional characterization was performed using scanning electron microscopy (SEM), transmission electron microscopy (TEM), and atomic force microscopy (AFM), enabling cross-validation of nanoscale feature dimensions and etch quality. Electrical transport measurements reveal periodic resistance minima and switching-current oscillations at magnetic fields set by the hole-array pitch, consistent with enhanced vortex pinning when the vortex configuration matches the patterned lattice. Cryogenic THz time-domain spectroscopy further reveals a reduced superfluid density in the nanopatterned region relative to pristine NbN, while both regions have a spectroscopic superconducting transition near 14~K. These results establish a robust fabrication platform for superconducting nanopatterns targeting vortex pinning, metamaterials, and quantum-device applications.

\end{abstract}


\maketitle

\section{Introduction}

\begin{figure*}[!htp]
\includegraphics{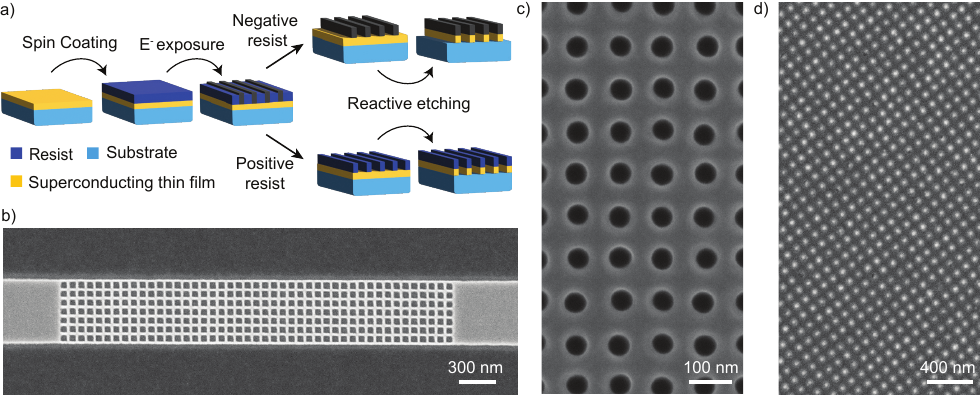}
\caption{\label{fig:1}Overview of the NbN nanopatterning processes and representative fabricated structures.
    (a) Schematic of the negative- and positive-tone electron-beam lithography processes, including resist spin coating, electron-beam exposure, development, and reactive-ion etching to transfer the pattern into the NbN film.
    (b) SEM micrograph of a 504-nm-wide nanostrip with a 70-nm-pitch grid-pattern, fabricated using the negative-tone HSQ process. The sample was coated with approximately 2~nm of Au for SEM imaging.
    (c) SEM micrograph of a 100-nm-pitch hole array patterned using the positive-tone ZEP~530A-6 process before resist removal.
    (d) SEM micrograph of the corresponding pattern transferred into the NbN film after resist stripping.}
\end{figure*}

Patterned superconducting thin films provide a versatile platform for quantum devices, superconducting electronics and detectors, engineered vortex pinning, and tailored optical and electromagnetic responses. Niobium nitride (NbN) is particularly attractive for these applications because it is a chemically and thermally robust refractory superconductor that retains a relatively high critical temperature at nanometre-scale thicknesses and can be patterned into nanoscale structures\cite{cucciniello_superconducting_2022}. It is used in superconducting photodetectors, high-frequency electronics \cite{colangelo_compact_2021,zhao_superconducting_2026}, and quantum devices \cite{liu_temperature-dependent_2017}. Epitaxial NbN is of particular interest for superconducting nanowire single-photon detectors (SNSPDs)
\cite{goltsman_picosecond_2001,esmaeil_zadeh_superconducting_2021}, where it can combine a high critical temperature with dirty-limit transport properties that favor efficient hotspot formation \cite{incalza_fast-recovery_2026}.

The lateral dimensions required for patterned superconductors depend on the application and can be as small as a few tens of nanometers. NbN nanobridge Josephson junctions, for example, may have widths of only a few tens of nanometers \cite{colletta_gap_2026}, whereas SNSPDs typically consist of nanowires tens to a few hundred nanometers wide \cite{incalza_fast-recovery_2026, marsili_single-photon_2011}. Similarly, artificial vortex-pinning landscapes employ periodic arrays of micro- or nanoscale holes, notches, or locally modified regions to control vortex motion, produce commensurability effects, and modify the field-dependent critical current \cite{bose_review_2023, thakur_vortex_2009}, and can also improve the magnetic-field resilience of superconducting devices \cite{kroll_magnetic-field-resilient_2019}.

More generally, reducing the pattern dimensions relative to the superconducting coherence length and magnetic penetration depth can qualitatively modify current flow and vortex dynamics. For example, highly ordered Nb nanomeshes with 34-nm periodicity have been shown to approach the theoretical depairing current density while retaining a critical temperature comparable to that of the continuous film \cite{xu_achieving_2010}. Dense periodic patterns are also relevant to nanophononic crystals, in which geometry is used to modify phonon dispersion and transport \cite{xie_brief_2023,ma_phonon_2022} and may influence superconducting properties \cite{mohammadi-arzanagh_enhanced_2025}. They are also relevant to electromagnetic metamaterials and metasurfaces, in which subwavelength structures provide an engineered electromagnetic response \cite{vafapour_sensing_2021,mirhosseini_superconducting_2018}.
Fabrication of such structures therefore requires lithographic processes that can define and transfer features at these length scales.

A variety of fabrication methods have been developed to define nanoscale structures in superconducting films. Optical lithography provides high throughput but generally lacks the resolution required for the smallest nanowires, constrictions, and dense pinning arrays. Focused-ion-beam milling enables direct-write patterning without a resist mask, but ion implantation, redeposition, and edge damage can locally suppress superconductivity \cite{tettamanzi_superconducting_2009}. Other approaches, including nanoimprint lithography \cite{sreenivasan_nanoimprint_2017,zhao_fabrication_2015}, self-assembled block copolymers \cite{jeong_directed_2013,robbins_block_2016}, colloidal masks, anodic-alumina templates, and superlattice-based nanowire pattern transfer \cite{xu_achieving_2010}, can produce dense nanoscale patterns over comparatively large areas. However, these methods generally offer less flexibility in pattern geometry, feature placement, and integration with contacts and alignment structures than direct-write electron-beam lithography (EBL). EBL therefore provides the resolution and pattern flexibility required for complex superconducting nanostructures, but translating this capability into reliable pattern transfer in ultrathin NbN remains challenging, especially when dense sub-100-nm features are targeted.

High-resolution EBL uses both negative- and positive-tone resists, with the choice depending on the target pattern. Negative-tone hydrogen silsesquioxane (HSQ) forms a silica-like network under electron irradiation and can serve as a high-resolution, etch-resistant mask. Optimized HSQ processing, including high-contrast salty development, has enabled dense patterns with sub-10-nm half pitch at the mask level \cite{yang_using_2007}. HSQ is also favorable for its stability during SEM imaging\cite{gangnaik_new_2017}. However, high-resolution patterning of ultrathin NbN presents additional challenges, particularly in maintaining HSQ adhesion during processing and subsequent pattern transfer. Positive-tone organic resists such as PMMA and ZEP are well suited to hole arrays because only the regions to be removed from the mask are exposed. ZEP 530A therefore provides a complementary approach for large-area arrays of nanoscale holes.

Here, we report two distinct top-down nanopatterning processes for epitaxial NbN films on sapphire. The first process, hereafter referred to as the HSQ process, was used to fabricate patterned nanostrip devices from approximately 5-nm-thick NbN films. This process employs HSQ-based electron-beam lithography optimized through development control, Si-assisted resist adhesion, and electron-beam proximity-effect correction. The resulting devices were characterized by scanning electron microscopy (SEM), transmission electron microscopy (TEM), and atomic force microscopy (AFM), and were used for electrical transport and magnetic-field-dependent measurements. The second process, hereafter referred to as the ZEP process, was used to define large-area hole-array regions in approximately 6-nm-thick NbN films. In this process, ZEP~530A was used as a positive-tone resist to pattern millimeter-scale periodic hole arrays. Cryogenic terahertz time-domain spectroscopy was then used to compare the temperature-dependent electromagnetic response and superfluid density of the large area hole-array regions. These fabrication routes enable studies of how periodic nanopatterning affects vortex transport and the THz electrodynamic response of NbN, providing a platform for superconducting metamaterial development.

\section{Experimental methods}
\label{sec:methods}

This section describes the fabrication of nanopatterned epitaxial NbN thin films and the experimental techniques used to characterize their structural, electrical, and superconducting properties.

\subsection{Nanofabrication}
\label{sec:fabrication}

The starting material consisted of epitaxial $\delta$-NbN thin films deposited by reactive DC magnetron sputtering on c-cut sapphire ($\mathrm{Al_2O_3}$) substrates\cite{incalza_fast-recovery_2026}. The sapphire wafers were 650~\textmu m thick and were diced into $10\times10~\mathrm{mm}^2$ chips prior to growth. Film growth was performed at a substrate temperature of 800\textdegree C\cite{incalza_fast-recovery_2026}.

Using the HSQ process, we fabricated 18 chips, denoted as HSQ1--HSQ18. The second used positive-tone ZEP~530A-6 to pattern millimeter-scale hole arrays. 6 chips were fabricated using this process and are denoted ZEP1--ZEP6. The NbN films on the A chips were 5-6~nm thick, whereas those on the B chips were 6-7~nm thick; film thicknesses were determined by ellipsometry. Most chips were used for process development and/or dedicated to destructive characterization. Of the HSQ process chips, HSQ6 was used for TEM imaging, HSQ10 for proximity effect correction testing and SEM characterization, HSQ11 for electrical characterization, and HSQ14 for magnetic field transport measurements. Of the ZEP process chips, ZEP4 was used for SEM characterization and ZEP6 was used for THz measurements.
In both processes, the developed resist served as an etch mask for pattern transfer into the NbN layer. The two fabrication routes are summarized schematically in Fig.~\ref{fig:1}(a), with representative structures shown in Figs.~\ref{fig:1}(b--d).

Electron-beam exposures were performed using an ELIONIX ELS-F125
system operated at an acceleration voltage of 125~kV. The pattern
generator was operated at 100~MHz, with a 500~\textmu m write field.
For the A chips, a 1-nm beam step size was used. In this case, the layouts occupied a lateral extent of approximately 17~\textmu m x 17~\textmu m and therefore fit entirely within a single 500 \textmu m write field, avoiding field-to-field stitching within the active patterned region. A beam current of 1~nA and a nominal exposure dose of 6400~$\text{ \textmu C}/\mathrm{cm}^2$ were used for these high-resolution nanowire structures. 

HSQ was diluted in methyl isobutylketone (MIBK) to concentrations of 1 and 2~wt.\%. The resist was spin-coated at 3000~rpm for 60~s. After development, the corresponding HSQ thicknesses were approximately 16~nm for the 1\% formulation and 32-33~nm for the 2\% formulation, as determined by AFM. The 2\% HSQ process was used as the baseline process for fabrication of the 70-nm-pitch patterned nanostrips. A standard development consisted of immersion in a 25\% aqueous solution of tetramethylammonium hydroxide (TMAH) for 2~min at room temperature, approximately 21\,\textdegree C, followed by rinsing in deionized (DI) water and drying with a nitrogen (N$_2$) gun.

Alternative development conditions were investigated to improve the resolution of the HSQ process. Motivated by previous studies showing that the development temperature can significantly affect HSQ contrast~\cite{chen_effects_2006, tatarintsev_effect_2020}, we evaluated the effect of developer temperature under our process conditions. Test structures consisting of single lines of EBL shots and line-to-line spacings ranging from 8 to 153~nm were developed in TMAH 25\% at either room temperature or 50\textdegree C.
A high-contrast salty-development process was also investigated using the 1\% HSQ formulation. In this case, the exposed samples were immersed for 30~s in an aqueous solution containing 4\% NaCl and 1\% NaOH at room temperature and subsequently rinsed in DI water, and dried with a with a N$_2$ gun. This chemistry follows the salty-development approach previously developed for high-resolution HSQ lithography~\cite{yang_using_2007, nam_contrast_2009}.

We also investigated whether a thin Si capping layer could improve the adhesion of thin HSQ films to epitaxial NbN. A nominally 2-nm-thick Si layer was deposited at room temperature without breaking vacuum after NbN growth, before HSQ coating. For the adhesion comparison, half of the NbN chip was masked during the Si deposition, allowing capped and uncapped regions to be patterned on the same substrate. Parallel-line structures with a 40-nm pitch were subsequently written in 1\% HSQ and developed for 2~min in 25\% TMAH at room temperature.

%For patterns processed with proximity-effect correction (PEC), this value served as the reference dose, while the local exposure dose was adjusted by the correction algorithm according to the surrounding pattern density. Exposure-dose series were also performed during process optimization to evaluate the sensitivity of the resulting critical dimensions to the applied dose.
\begin{figure*}
\includegraphics{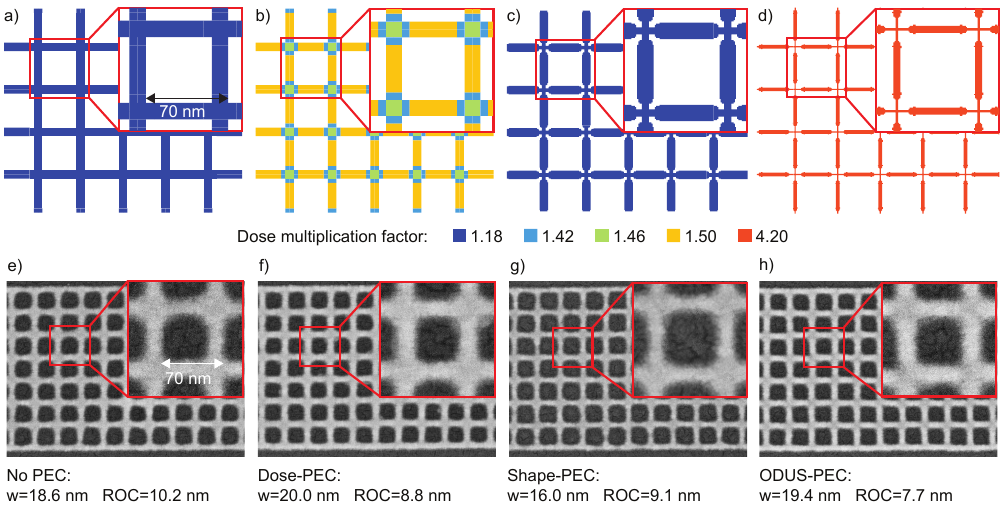}
\caption{\label{fig:3}Comparison of proximity-effect-correction (PEC) strategies for nanowire grids with a 70~nm pitch.
    (a) Nanowire layout without short-range PEC; the inset shows a magnified view of a square unit cell.
    (b) Layout with dose-based PEC and a minimum feature size (MFS) of $7 \times 7$~nm$^2$.
    (c) Layout with shape-based PEC without overdosing.
    (d) Undersized shape-PEC layout with a threefold exposure-dose factor, labeled as over-dose/unsder-size (ODUS) PEC.
    (e--h) Corresponding SEM micrographs after development and pattern transfer by etching for the layouts shown in (a--d), respectively. The average subwire width (w) and the minimum radius of curvature (ROC) of the hole corners are indicated. Before SEM imaging, the samples were coated with approximately 2~nm of Au to reduce charging-related artifacts.}
\end{figure*}
Proximity-effect correction (PEC) was implemented using GenISys BEAMER. The electron-scattering point-spread function used for the correction was estimated using Monte Carlo simulations in GenISys TRACER, with the short-range contribution represented using a 10-nm full width at half maximum gaussian profile. PEC tests were performed on a grid geometry consisting of nominally 56-nm-wide square openings separated by 14-nm-wide NbN regions, corresponding to a 70-nm pattern pitch. Several correction strategies were compared, including dose-based PEC with a minimum fracturing size of 7~nm, shape-based correction without intentional overdosing\cite{ocola_nanoscale_2009}, and combined over-dose/under-size (ODUS) correction using overdose factors of 3\cite{gorelick_direct_2010}.
Uncorrected devices and devices fabricated using the different PEC strategies were written together in chip HSQ10 as dose series to enable direct comparison under otherwise identical processing conditions.

For the HSQ process, after development, the resist pattern was transferred into the NbN
film using a $\mathrm{CF_4}$-based reactive-ion-etching process in a
PlasmaTherm Corial 210RL RIE system. The etch recipe had previously been developed for NbN processing and was used here without further optimization. For approximately 5-nm-thick NbN films patterned with this process, the samples were etched for approximately 30~s with a chamber pressure of 10 mTorr.

For the nanopatterning of mm$^2$ areas, ZEP~530A-6 was used as a positive-tone electron-beam resist on approximately 6-nm-thick NbN films. The resist was spin-coated at 5000~rpm for 60~s and baked at 180\textdegree C for 120~s. Periodic arrays of single-spot exposures were then written at a beam current of 20~nA with a nominal pitch of 100~nm. Conventional point-by-point writing, in which the beam was blanked between neighboring exposure points, was used for the large-area patterns, with a dwell time of 1.02~\textmu s per point and no PEC.
Of the fabricated samples, ZEP4 had a 6.2 nm thick NbN film, and contained a $2~\mathrm{mm}\times1~\mathrm{mm}$ patterned region assembled from $500~\text{\textmu}\mathrm{m}\times500~\text{\textmu}\mathrm{m}$ EBL write fields and is shown in Fig.~4. Chip ZEP6 had a 6.5 nm thick NbN film, and a total patterned area of $2~\mathrm{mm}\times2~\mathrm{mm}$ and was used for the THz measurements. Neighboring write fields were separated by a nominally 300-nm-wide unpatterned stitching region to prevent overlapping holes that would electrically open the pattern, as shown in Fig.~4(b).
For the ZEP process, the resist was developed for 60~s
in \textit{o}-Xylene at 0\textdegree C, rinsed in isopropyl alcohol for 30~s at room temperature, and dried with a N$_2$ gun. The resulting pattern was transferred into the NbN using the same $\mathrm{CF_4}$-based RIE process, with an etch time of 45~s. Prior to characterrization, the resist was stripped in N-methyl-2-pyrrolidone at 70\,\textdegree C for 4 hours.

\subsection{Structural and dimensional characterization}
\label{sec:characterization}

Scanning electron microscopy (SEM) was used throughout process development to evaluate resist-mask quality and pattern fidelity. SEM imaging was performed using a Zeiss Gemini 450. 
For the HSQ process, samples imaged immediately after HSQ development were examined at an accelerating voltage of 1~kV and a beam current of 100~pA. During these measurements, the thin NbN film was electrically grounded using metallic clips to reduce charging artifacts. 

When imaging nanowires after the NbN etching step, charging from the exposed sapphire substrate became substantially more pronounced. Samples intended for high-resolution post-etch SEM imaging were therefore coated with approximately 2~nm of sputtered Au. These samples were imaged using an accelerating voltage of 3~kV, a beam current of 100~pA, and an in-lens detector. The Au coating was applied only to samples for which electrical characterization was not planned. For the positive tone process, SEM was used to examine the sample before and after resist stripping, with a current of 100~pA and accelerating voltage of 5~kV and 1.5~kV respectively.

Transmission electron microscopy (TEM) was used to evaluate the cross-sectional structure of the nanopatterned devices and to verify transfer of the lithographic pattern to the NbN layer.
TEM lamellae were prepared from chip HSQ6 using a FEI Helios Nanolab 600 DualBeam SEM--focused-ion-beam system equipped with a Ga ion source and an
Omniprobe micromanipulator. 
To process the sample on the insulating sapphire substrate, a 3~nm Au layer was sputter-deposited to reduce charging. %Prior to FIB milling, a protective Pt layer was deposited over the region of interest, first by electron-beam-induced deposition and subsequently by ion-beam-induced deposition. 
To directly observe the hole morphology in cross-sectional TEM, the lamella must intersect a row of holes and have a thickness smaller than the hole width. However, alignment of the lamella with an individual hole row was made difficult by charging of the sapphire substrate. To improve the probability of obtaining an aligned cross section, multiple nominally identical nanowire structures were fabricated next to one another, each with a small lateral offset. A single lamella was then prepared across the array so that at least one nanowire was aligned with the cross-sectional plane.
Imaging was subsequently performed using a FEI Tecnai Multipurpose Digital TEM. Feature dimensions were measured from the TEM images using ImageJ.

AFM provided a complementary, non-destructive measurement of the nanopattern topography. Measurements were performed using an Asylum Research Jupiter XR AFM. High-resolution scans were acquired in tapping mode using Bruker TESPA-V2-SS probes. %which have a nominal tip radius of 2-5~nm and a half-angle below 10$^\circ$ over the first approximately 200~nm of the tip 
Regions of interest were first located using non-contact imaging before switching to a low-force tapping regime. A new probe was used for each imaging session to minimize the effects of tip wear and contamination. Because the measured lateral dimensions are affected by convolution between the AFM probe geometry and the nanoscale features, SEM and cross-sectional TEM were used to provide complementary information on the lateral dimensions.

\subsection{Electrical transport and THz measurements}
\label{sec:transport_thz}

Electrical transport measurements were performed on nanopatterned strip devices and on reference NbN structures without periodic holes, all fabricated on chip HSQ11.
The resistance was measured as a function of temperature across the
superconducting transition using a four-terminal configuration. The
fitted $T_\mathrm{C}$ were subsequently used to compare patterned devices with the corresponding unpatterned reference structures. To compare devices using a consistent resistive-transition criterion,
each measured $R(T)$ curve was fitted by least squares to an empirical four-parameter logistic function,
$R(T) = R_0 + \frac{\Delta R}{1+\exp[-(T-T_C)/w]}$.
Here, $R_0$ and $R_0+\Delta R$ are the fitted low- and high-temperature resistance levels, respectively, and $w$ is a width parameter. We define $T_C$ as the midpoint of the fitted transition,
$R(T_C)=R_0+\Delta R/2$. The fit was used to estimate this transition midpoint for comparison between devices.
Current-voltage characteristics were also measured below the superconducting transition. Positive- and negative-bias sweeps were analyzed separately to extract the
corresponding switching currents, $I_{\mathrm{sw}}^{+}$ and $I_{\mathrm{sw}}^{-}$.

Magnetic-field-dependent transport measurements were performed on a nanopatterned strip device fabricated on chip HSQ14. An identical device on the same chip is shown in the AFM image in Fig.~\ref{fig:7}(b). Measurements were performed using a Quantum Design MPMS system with the magnetic field applied perpendicular to the chip. For the resistance measurements, an AC current of 100~nA at 0.436~Hz and an averaging time of 4.59~s were used. For the switching-current measurements, the applied current was repeatedly swept between $-60$ and $60~$\textmu A.
Resistance was measured as a function of applied magnetic field at several fixed temperatures near the superconducting transition, and switching-current measurements were performed as a function of field at 4~K. 

On sample ZEP6, fabricated with the ZEP process, we performed cryogenic terahertz time-domain spectroscopy (THz-TDS) using a commercially available  Menlo Systems photoconductive-antenna (PCA) emitter–detector pair. The laser operates at 80 MHz  repetition rate, covers 0–6 THz, with a dynamic range greater than 40 dB over 0.2–3 THz. Using this cryogenic THz system, we measured the amplitude and phase of the THz transmission through pristine and nanopatterned regions of the same 6.5-nm-thick superconducting NbN film as a function of temperature across the superconducting transition.

\section{Results and discussion}

This section first presents optimization of the HSQ process used to fabricate patterned nanostrip devices and then describes the ZEP process used to fabricate large-area hole-array regions. It subsequently presents structural characterization of the patterned nanostrip devices, superconducting transport measurements of these devices, and THz characterization of the large-area hole-array regions.

\subsection{Optimization of high-resolution HSQ patterning}
\label{sec:results_hsq}

The negative-tone 2\% HSQ process with 25\% TMAH development at room temperature yielded reproducible fabrication of 70-nm-pitch grids in nanostrip devices, while attempts to increase the pattern density resulted in reduced pattern fidelity and incomplete development and etching of the holes. Fig.~\ref{fig:1}(b) displays an example nanostrip device fabricated with this process on chip HSQ10, with a total width of 504 nm, a patterned length of 3.14~\textmu m, and a pattern pitch of 70~nm.

% \begin{figure}[h]
%     \centering
%     \includegraphics[width=\linewidth]{figures/Asset 2.pdf}
%     \caption{SEM micrographs of HSQ test structures patterned on NbN under different adhesion and development conditions. (a) HSQ lines with a 40~nm pitch patterned on uncapped NbN and (b) on NbN capped with a 2-nm-thick Si adhesion layer. (c) Test grids patterned in 2\% HSQ on a 5-nm-thick NbN film on sapphire and developed at room temperature and (d) at 50~\textdegree C. The grids comprise perpendicular single-shot lines with nominal spacings from 8 to 153~nm. (e) HSQ mask developed using the salty-development process. Although limited pattern fidelity and local adhesion failure are observed, a pitch of 30~nm is achieved.}
% \label{fig:2}
% \end{figure}

When patterning dense HSQ features directly on epitaxial NbN, delamination was frequently observed after development. Similar substrate-dependent adhesion limitations have been reported for HSQ on a range of metallic and dielectric surfaces, with adhesion strongly influenced by surface chemistry \cite{zhang_improving_2014}. Introducing a 2~nm  layer of sputtered Si substantially improved adhesion and preserved the patterned features after development. The improved adhesion observed after introducing the Si capping layer is consistent with previous reports showing that thin Si interlayers can promote HSQ adhesion\cite{radtke_reliable_2019}, likely through the formation of a native SiO$_2$ surface that provides favorable Si--O--Si bonding with the exposed HSQ resist.

% The effect of HSQ development temperature was examined by developing exposed HSQ in 25\% TMAH at 21\,\textdegree C and at 50\,\textdegree C. Because development temperature also changes HSQ sensitivity, a separate dose test was performed for each temperature, so that each condition was evaluated at its own dose range. At 50\,\textdegree C, the transition between densely and less densely patterned regions is more uniform than at 21\,\textdegree C, which we attribute to increased development contrast.

% Delamination was observed for patterns developed at 50\,\textdegree C, whereas those developed at room temperature remained attached. Although elevated-temperature TMAH development is commonly used to improve HSQ development contrast\cite{tatarintsev_effect_2020, chen2006 }, it may also promote failure at a weak HSQ/substrate interface.

The use of a salty developer enabled a further reduction in pattern pitch compared with the 25\% TMAH process. 30-nm-pitch structures were successfully resolved using 1\% HSQ developed for 30~s in a 4\% NaCl--1\% NaOH solution after optimization of the exposure dose.
However, the resulting mask still exhibited substantial distortion, local delamination, and edge defects.

The effect of different PEC strategies is compared in Fig.~\ref{fig:3}.

Without short-range PEC, the intersections between perpendicular lines are locally overexposed [Fig.~\ref{fig:3}(a,e)], leading to rounded corners and reduced fidelity of the nominally square openings. Dose-based PEC with a 7~nm minimum feature size improves the pattern definition by locally adjusting the exposure dose [Fig.~\ref{fig:3}(b,f)], while shape-based PEC compensates for the excess exposure at the intersections by modifying the layout geometry [Fig.~\ref{fig:3}(c,g)]. The over-dose under-size approach \cite{gorelick_direct_2010}, in which the corrected features are intentionally undersized and exposed at a higher dose, also produces well-defined square openings [Fig.~\ref{fig:3}(d,h)].
To quantify corner rounding as a function of PEC, we analyzed approximately 100 square-hole openings for each device in Fig.~\ref{fig:3}. Before analysis, the SEM micrographs were smoothed with a 2-pixel median filter and automatically thresholded in ImageJ to produce binary images. For each opening, we extracted a fixed 70 nm × 70 nm crop from the binary image and analyzed its contour using ImageJ’s Contour Curvature plugin with 8 coefficient pairs. We selected the four strongest positive curvature maxima, and the mean of these four peak curvatures was used as the corner-curvature metric.
The minimum corner radius resulted 10.2 nm for Fig.~\ref{fig:3}(e), 8.8 nm for Fig.~\ref{fig:3}(f), 9.1 nm for Fig.~\ref{fig:3}(g), and 7.7 nm for Fig.~\ref{fig:3}(h). 
The dose-based and over-dose under-size approaches both improve the corner sharpness, with the latter showing the best result. 

The smoothed and thresholded images were also used to characterize the width of the NbN region separating adjacent square holes, hereafter referred to as a subwire. The average subwire width was measured as 18.6 nm for Fig.~\ref{fig:3}(e), 20.0 nm for Fig.~\ref{fig:3}f), 16.0 nm for Fig.~\ref{fig:3}(g), and 19.4 nm for Fig.~\ref{fig:3}(h). The narrower subwires obtained with shape-based PEC are attributed to the reduced total written area introduced by the geometry correction, which was not compensated by a corresponding dose correction.

\subsection{Large-area nanopatterning using ZEP 530A-6}
\label{sec:results_zep}

\begin{figure}[h!]
    \centering
    \includegraphics[width=\linewidth]{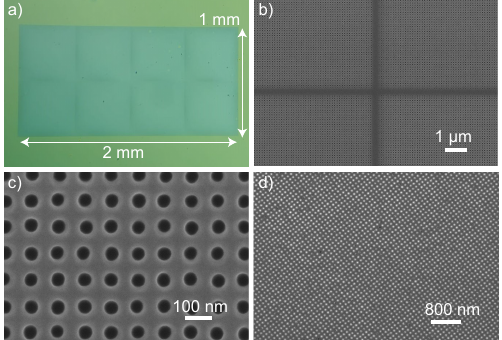}
    \caption{Patterning of a square array of holes with a 100~nm pitch over an area of 2~mm$^2$.
    (a) Optical micrograph of the patterned area after development.
    (b) SEM micrograph acquired at the junction of four electron-beam write fields.
    (c) SEM micrograph of a portion of the patterned area before resist stripping.
    (d) SEM micrograph of the pattern after resist stripping, showing substrate charging in the hole regions.}
\label{fig:4}
\end{figure}

The positive-tone ZEP process allowed periodic nanopatterning to be extended from individual nanostrip devices to millimeter-scale NbN films. Figure~\ref{fig:4}(a) shows a representative $2~\mathrm{mm}\times1~\mathrm{mm}$ patterned region.

The main nonuniformity appeared near the boundaries between adjacent EBL write fields. As shown in Fig.~\ref{fig:4}(b), the field boundaries are visible, and the holes closest to these regions appear smaller than those farther from the boundaries. Away from the field edges, the periodic array is substantially more uniform, as shown in Fig.~\ref{fig:3}(c). The hole diameter in the mask in these regions was measured to be approximately 47--52~nm for a nominal pitch of 100~nm.

After reactive-ion etching and removal of the ZEP mask, the periodic hole array remains clearly visible in the NbN film [Fig.~\ref{fig:4}(d)]. The hole diameter in the NbN film is approximately 30~nm.

This confirms that the positive-tone pattern was transferred into the NbN film. The hole diameter appears about 40\% smaller after resist stripping than in the top-down SEM of the resist. Compared with the HSQ process, the ZEP process enables substantially larger patterned areas, at the expense of minimum achievable feature size and control over the feature shape.The higher sensitivity of ZEP reduces the required exposure dose, while the positive-tone geometry defines each hole through a single dot exposure rather than by writing the surrounding resist network. This simpler geometry also permits to use a larger beam current. These throughput advantages motivated the adoption of the ZEP-based process for millimeter-scale hole arrays.

\subsection{Dimensional validation and pattern transfer}
\label{sec:results_metrology}

\begin{figure}[h!]
    \centering
    \includegraphics[width=\linewidth]{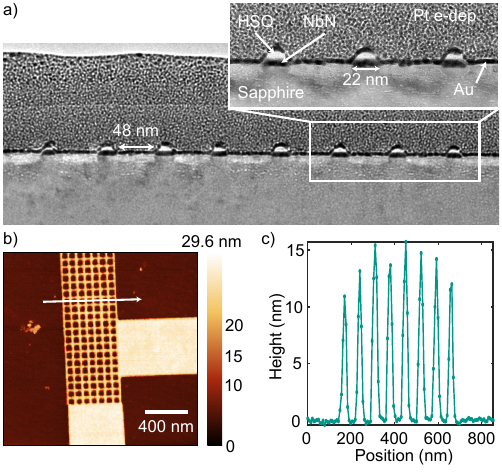}
    \caption{Characterization of a 500-nm-wide patterned nanostrip with a square hole array of 70-nm pitch. (a) Cross-sectional TEM image across a row of holes. The holes have a cross-sectional dimension of 48~nm. The inset shows a magnified view of three sub-wires, each with a lateral dimension of 22~nm. The HSQ mask is the brightest region, with NbN visible beneath it. The Au coating appears dark and covers the hole regions and the HSQ mask. An electron-beam-deposited Pt protective layer is also visible above the patterned region. (b) Tapping-mode AFM height map of one of the nanostrips. Two of the four measurement leads are visible in the image. The white arrow marks the line across a row of holes along which the height profile in (c) was acquired. (c) Corresponding AFM height profile.}
\label{fig:7}
\end{figure}

Cross-sectional TEM provided direct confirmation that the periodic pattern was transferred through the NbN layer [Fig.~\ref{fig:7}(a)].

For a representative structure on chip HSQ6 with a 70 nm pitch, the NbN region separating adjacent square holes, hereafter referred to as a subwire, was approximately 22 nm wide. The square holes were approximately 48 nm wide. The different layers are distinguishable by their contrast: the bright regions correspond to the remaining HSQ mask, the darker regions beneath it to NbN, and the Au coating covers the structure and exposed sapphire. The granular layer at the top corresponds to Pt deposited during lamella preparation. The cross section confirms removal of the NbN between adjacent subwires, while the rounded HSQ profile indicates sloping of the mask sidewalls during processing. Some modification of the sapphire surface is also visible in the etched regions, indicating a range of overetch into the substrate.

AFM provided complementary, non-destructive characterization of structures fabricated using the same process conditions. Figure~\ref{fig:7}(b) shows a representative topographic map of a 70-nm-pitch patterned nanostrip, in which the periodic height modulation is resolved. The corresponding line profile in Fig.~\ref{fig:7}(c) shows a regular sequence of topographic peaks with heights of approximately 11-16~nm. The measured topographic height is lower than the initial combined thickness of the HSQ and NbN layers, with approximately 6--11~nm of HSQ estimated to remain as the masking layer, consistent with the TEM results. 
%Notably, the topographic height measured at the center of the patterned holes is comparable to that of the surrounding unpatterned regions, consistent with successful etching of the NbN layer within the holes.

The agreement between the TEM and AFM measurements provides complementary evidence of successful fabrication. TEM directly resolves the etched NbN cross section and therefore provides the most reliable information on the local feature geometry, whereas AFM provides non-destructive confirmation of the periodic surface topography. Lateral dimensions extracted from AFM should nevertheless be interpreted with caution because of tip-sample convolution, particularly for features with widths of only a few tens of nanometers.

\subsection{Superconducting transport and field-dependent response}
\label{sec:results_transport}
\begin{figure}[h!]
    \centering
    \includegraphics[width=\columnwidth]{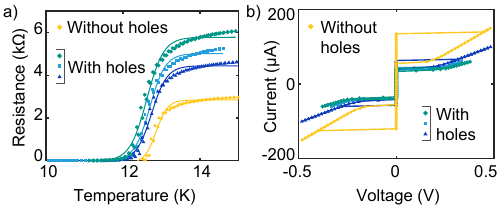}
    \caption{Electrical measurements of 500-nm-wide nanowires with and without 70-nm-pitch square hole arrays. Devices with holes were designed to be identical.(a) Resistance as a function of temperature for three patterned nanowires and one unpatterned nanowire. (b) Current--voltage characteristics of the same devices, showing reduced switching currents and hysteresis in the patterned nanowires. Measurements were performed at 3.2~K using a four-probe configuration.}
\label{fig:5}
\end{figure}

Electrical transport measurements were used to assess whether the HSQ nanopatterning process preserved superconducting behavior and to investigate the response of the patterned structures to an applied magnetic field. Fig.~\ref{fig:5}(a) compares the resistance--temperature characteristics of nanostrips patterned with square arrays of holes having a 70~nm pitch with those of an unpatterned reference.

All measured devices exhibited a superconducting transition within the same temperature range, indicating that superconductivity is retained after
electron-beam lithography and pattern transfer. The patterned devices
showed a higher normal-state resistance, consistent with the reduction
of the effective conducting cross section introduced by the periodic
array of holes. Despite having nominally identical hole and nanostrip-width dimensions and being fabricated in the same chip, the patterned devices exhibited some device-to-device
variation in both $T_\mathrm{C}$ and switching current. This variation
may arise from small local differences in feature dimensions, edge
roughness, film properties, or current distribution that are not
captured by the nominal design dimensions.

The solid curves in Fig.~\ref{fig:5}(a) correspond to the logistic fits described in Sec.~\ref{sec:transport_thz}. The extracted transition temperatures are \(T_\mathrm{C}=12.7\), \(12.8\), and \(12.6~\mathrm{K}\) for the three patterned devices, respectively, compared with \(T_\mathrm{C}=12.865~\mathrm{K}\) for the unpatterned reference. Thus, the transition temperatures remain close to that of the unpatterned reference, showing that the fabrication and patterning processes did not produce a pronounced suppression of \(T_\mathrm{C}\). %This is particularly relevant because the periodic pattern substantially modifies the local geometry of the nanowire while leaving the superconducting transition largely preserved.

The current-voltage characteristics provide a complementary comparison of the patterned and unpatterned devices [Fig.~\ref{fig:5}(b)]. Both retained a well-defined zero-resistance branch followed by a transition to the resistive state. The unpatterned reference exhibited a switching current of 135 \textmu A, whereas the three patterned devices switched at 63, 44, and 42~\textmu A, respectively. The reduced switching currents of the
patterned devices were consistent with the reduced current-carrying cross
section and the redistribution of current around the holes. %The reduced switching current therefore accompanies the geometric modification of the nanowire without eliminating its superconducting state.

\begin{figure}[h!]
    \centering
    \includegraphics[width=\linewidth]{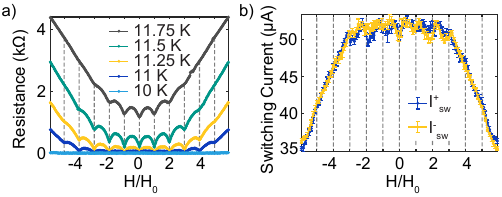}
\caption{Magneto-transport measurements of a 500-nm-wide nanostrip with 70-nm-pitch square hole arrays. (a) Resistance versus applied perpendicular field, in units of the matching field $H_0=4220$~Oe, measured at temperatures near the film's transition temperature ($T_\text{C}$ = 12.2 K). (b) Switching current measurements of the same nanowire at 4 K. Each data point represents the average and standard deviation of 15 measurements.}
\label{fig:6}
\end{figure}

The magnetic-field transport measurements are shown in Fig. \ref{fig:6}. 
These measurements were performed on a different sample from that measured in Fig. \ref{fig:5}, but with a similar geometry with 70 nm pitch grid nanostructures, a width of 504 nm, and a length of 3.14 \textmu m. This device has a NbN film thickness of 6.2 nm, a critical temperature of $T_\mathrm{C}=12.2$\,K and a switching current of $I_\mathrm{sw}=50$\,\textmu A.
For these measurements, the magnetic-field axis is normalized to the characteristic matching field $H_0$, given by
\begin{equation}
H_0 = \frac{\Phi_\circ}{\mu_0 a^2},
\label{eq:Matching_Field}
\end{equation}

\noindent where $\mu_0$ is the permeability of free space, $\Phi_\circ = \frac{h}{2e}$ is the flux quantum, and $a^2$ is the area of the hole-lattice cell \cite{bose_review_2023}. Using $a = 70$~nm, we obtain $H_0 = 4220$~Oe.

In the resistance vs field measurements in Fig.~\ref{fig:6}(a), a distinct signature of the vortex matching effect is visible, with resistance minima at multiple values of $H_0$, visible up to $H/H_0=5$ for $T=$11~K. The observation of multiple matching fields provides evidence of commensurate vortex pinning by the periodic hole array in the NbN nanowire. %The features become most pronounced in the temperature range where the device is partially resistive and therefore particularly sensitive to field-induced vortex motion.

Figure~\ref{fig:6}(b) shows the magnetic-field dependence of the absolute value of the switching current, measured for both current polarities in the same patterned nanowire. For fields up to $|H/H_0|=3$, the switching current was weakly field-dependent, non-monotonic, and exhibited oscillations with a period corresponding to one matching field.
Notably, at $H=3H_0=12660~\mathrm{Oe}=1.266$~T, the switching current was comparable to its zero-field value.
The approximately periodic response in both resistance and switching current was consistent with a commensurability effect between the vortex configuration and the artificial hole lattice.

Taken together, these measurements show that the HSQ nanopatterning process preserved the superconducting transition while introducing a geometry-dependent magnetic-field response. The periodic modulation in magnetoresistance and the broad region of weak field dependence in the switching current provide evidence that the nanoscale hole array acts as an engineered vortex-pinning landscape in the patterned nanowire.

\subsection{THz measurements in nanopatterned films}
To independently probe the superconducting state in the pristine and nanopatterned regions, we performed cryogenic terahertz time-domain spectroscopy (THz-TDS) using the system described in Section~\ref{sec:transport_thz}. Whereas electrical transport characterizes the zero-resistance state, THz-TDS probes the inductive response of the superconducting condensate. This response provides access to the superfluid stiffness $\rho_s$, which determines the strength of the superconducting screening response and is proportional to the inverse square of the London penetration depth.

\begin{figure}[h!]
    \centering
    \includegraphics[width=\linewidth]{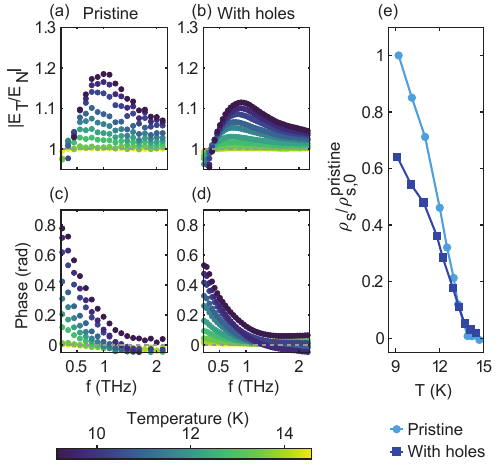}
    \caption{Temperature-dependent THz transmission of pristine and nanopatterned regions of a 6.5-nm-thick NbN film. (a,b) Magnitude and (c,d) phase of the transmitted field at each temperature normalized to the normal-state reference field. Panels (a,c) show the pristine region, and panels (b,d) show the region with hole arrays. (e) Superfluid stiffness normalized to the value of the pristine region at the lowest measured temperature.}
\label{fig:8}
\end{figure}

We measured the amplitude and phase of the THz transmission through pristine and nanopatterned regions of the same 6.5-nm-thick superconducting NbN film as a function of temperature across the superconducting transition. The film had a patterned area of 2 mm $\times$ 2 mm and a square array of holes with a 100 nm pitch and a diameter of approximately 30 nm when measured in a representative region in the center of a write field. Fig.~\ref{fig:8} shows the amplitude and phase of the transmitted THz field at each temperature, normalized to the normal-state spectrum of the same region, \(r = E_T/E_N\).

Both the amplitude [Fig. \ref{fig:8}(a,b)] 
and phase [Fig. \ref{fig:8}(c,d)] change markedly upon cooling through the transition. The amplitude develops a 
rapid increase near 0.5 THz and a broad maximum near 1 THz, while the low-frequency phase response 
grows with decreasing temperature. The onset of these two spectroscopic responses occurs near 14 K, indicating a superconducting transition temperature of \(T_c \approx 14~\mathrm{K}\) in both regions. For a thin film, the maximum of \(|E_T/E_N|\) occurs at the gap frequency \(2\Delta\) where pair-breaking absorption sets in\cite{glover_conductivity_1957}. The maxima observed near 1 THz  in both regions thus correspond to \(2\Delta\approx4~\mathrm{meV}\) at 9 K, in line with the theoretical BCS value \(3.5\,k_{\mathrm{B}}T_c\approx4~\mathrm{meV}\) for \(T_c\approx14~\mathrm{K}\). The maxima of the two regions differ by no more than about 15\%, within the frequency resolution of the pristine data, so the patterning leaves the gap scale at most modestly changed, while, as shown below, pattering apparently reduces the superfluid stiffness substantially. From the complex transmission, we extracted the normalized superfluid stiffness as a function of temperature [Fig. \ref{fig:8}(e)]. 

For a film much thinner than the wavelength, the ratio of the fields transmitted in the superconducting and normal states is \(r = (1+n+Z_0G_N)/(1+n+Z_0G_s)\), where \(n\) is the substrate index, \(Z_0\) the impedance of free space, and \(G_s = G_1 + iG_2\) the complex sheet conductance of the film \cite{glover_conductivity_1957,nuss_dynamic_1991}; because THz-TDS records amplitude and phase, \(G_1\) and \(G_2\) follow without Kramers--Kronig analysis \cite{nuss_dynamic_1991}. The condensate contributes the inductive part, \(\omega G_2(\omega\to0) = n_s d\,e^2/m = d/\mu_0\lambda^2\)\, \cite{nuss_dynamic_1991}, so the superfluid stiffness \(\rho_s = \hbar^2 n_s d/4m\) is obtained directly as \(\rho_s \propto \omega\,\mathrm{Im}[1/r(\omega)]\), averaged over 0.3--0.6 THz where this product is frequency independent. Fig.~\ref{fig:8}(e) shows \(\rho_s(T)\) of both regions divided by the pristine value at the lowest temperature (9.2 K); the temperature-independent prefactor \(1+n+Z_0G_N\) cancels within each region and differs by 5\% between them. At the lowest measured temperatures, the superfluid stiffness of the nanopatterned region is approximately \(62\%\) of the pristine value. This reduction is far larger than the estimated area reduction from the nanopatterned holes. %Attributed entirely to a non-superconducting depletion region around each hole, it would correspond to a rim about 20 nm wide beyond the 30-nm hole edge, whereas a weaker, spatially uniform reduction of the stiffness in the remaining film would give the same average depletion. The present data do not distinguish between these limits. In either case the hole array lowers the superfluid stiffness, by a factor of about 1.6 rather than the 1.07 expected from the removed area alone.

\label{sec:results_THz}

\section{Conclusions}

We demonstrated a top-down fabrication process for nanopatterning ultrathin epitaxial NbN films on sapphire. By combining high-resolution electron-beam lithography, optimized development processes, adhesion-layer integration, proximity-effect correction, and multimodal characterization, we established a process platform for periodically patterning superconducting nanowires and large-area NbN films. In particular, the observed vortex commensurability effects demonstrate that the periodic hole array controls the magnetic-field dependence of superconducting transport, with a characteristic field scale set by the nanoscale lattice pitch. These results provide a route toward superconducting metamaterial devices, engineered vortex pinning landscapes, and nanoscale quantum devices based on epitaxial NbN.

\begin{acknowledgments}
We wish to acknowledge the support of MIT.nano for providing access to nanofabrication and characterization facilities. We also thank J. Daley and C. Moorman for their assistance with SEM imaging, R. Dana and J. Piety for help with AFM characterization, Z. Zhang for assistance with TEM lamella preparation and imaging, M. Mondol for help with e-beam lithography, and J. Pettine for help with device fabrication. We are also grateful to Rohit Prasankumar, Matthew Julian, Stu Wolf, Quanxi Jia, and Mukund Vengalattore for helpful discussions, as well as to Phillip Donald Keathley, Reed Foster, and Malick Sere for their insightful feedback on this manuscript. DM acknowledges support from the Alan L. McWhorter fellowship, DJP acknowledges support from the MathWorks Engineering Fellowship, and AS acknowledges support from the NSF Graduate Research Fellowship Program (GRFP). This work was funded in part by the Defense Sciences Office (DSO) of the Defense Advanced Research Projects Agency (DARPA) (HR0011-24-9-0311). Research at Columbia was supported by ONR-N000142612225. The final stages of the project, including preparation of this manuscript, were supported by the National Science Foundation under Award No. 2617411.

\end{acknowledgments}

\section*{Data Availability}
The data that support the findings of this study are available from the corresponding author upon reasonable request.

\bibliography{references_v2}

@article{incalza_fast-recovery_2026,
	title = {Fast-{Recovery} {Epitaxial} {NbN} {Superconducting} {Nanowire} {Single}-{Photon} {Detectors} with {Saturated} {Efficiency} at 1550 nm in {Liquid} {Helium}},
	volume = {26},
	issn = {1530-6984},
	url = {https://doi.org/10.1021/acs.nanolett.6c00261},
	doi = {10.1021/acs.nanolett.6c00261},
	number = {21},
	urldate = {2026-09-18},
	journal = {Nano Letters},
	author = {Incalza, Francesca and Castellani, Matteo and Joti Paul, Dip and Simon, Alejandro and Batson, Emma and Mondin, Davide and Medeiros, Owen and Berggren, Karl K.},
	month = jun,
	year = {2026},
	pages = {6848--6856},
}

@article{thakur_vortex_2009,
	title = {Vortex matching effect in engineered thin films of {NbN}},
	volume = {94},
	issn = {0003-6951},
	url = {https://doi.org/10.1063/1.3167771},
	doi = {10.1063/1.3167771},
	number = {26},
	urldate = {2026-03-30},
	journal = {Applied Physics Letters},
	author = {Thakur, Ajay D. and Ooi, Shuuichi and Chockalingam, Subbaiah P. and Jesudasan, John and Raychaudhuri, Pratap and Hirata, Kazuto},
	month = jun,
	year = {2009},
	pages = {262501},
}

@article{esmaeil_zadeh_superconducting_2021,
	title = {Superconducting nanowire single-photon detectors: {A} perspective on evolution, state-of-the-art, future developments, and applications},
	volume = {118},
	issn = {0003-6951},
	shorttitle = {Superconducting nanowire single-photon detectors},
	url = {https://doi.org/10.1063/5.0045990},
	doi = {10.1063/5.0045990},
	number = {19},
	urldate = {2025-09-06},
	journal = {Applied Physics Letters},
	author = {Esmaeil Zadeh, Iman and Chang, J. and Los, Johannes W. N. and Gyger, Samuel and Elshaari, Ali W. and Steinhauer, Stephan and Dorenbos, Sander N. and Zwiller, Val},
	month = may,
	year = {2021},
	pages = {190502},
}

@article{goltsman_picosecond_2001,
	title = {Picosecond superconducting single-photon optical detector},
	volume = {79},
	issn = {0003-6951, 1077-3118},
	url = {https://pubs.aip.org/apl/article/79/6/705/515874/Picosecond-superconducting-single-photon-optical},
	doi = {10.1063/1.1388868},
	number = {6},
	urldate = {2026-09-21},
	journal = {Applied Physics Letters},
	author = {Gol’tsman, G. N. and Okunev, O. and Chulkova, G. and Lipatov, A. and Semenov, A. and Smirnov, K. and Voronov, B. and Dzardanov, A. and Williams, C. and Sobolewski, Roman},
	month = aug,
	year = {2001},
	pages = {705--707},
}

@article{cucciniello_superconducting_2022,
	title = {Superconducting niobium nitride: a perspective from processing, microstructure, and superconducting property for single photon detectors},
	volume = {34},
	issn = {0953-8984, 1361-648X},
	shorttitle = {Superconducting niobium nitride},
	url = {https://iopscience.iop.org/article/10.1088/1361-648X/ac7dd6},
	doi = {10.1088/1361-648X/ac7dd6},
	number = {37},
	urldate = {2026-09-21},
	journal = {Journal of Physics: Condensed Matter},
	author = {Cucciniello, Nicholas and Lee, Derek and Feng, Henry Y and Yang, Zihao and Zeng, Hao and Patibandla, Nag and Zhu, Mingwei and Jia, Quanxi},
	month = sep,
	year = {2022},
	pages = {374003},
}

@article{colangelo_compact_2021,
	title = {Compact and {Tunable} {Forward} {Coupler} {Based} on {High}-{Impedance} {Superconducting} {Nanowires}},
	volume = {15},
	issn = {2331-7019},
	url = {https://link.aps.org/doi/10.1103/PhysRevApplied.15.024064},
	doi = {10.1103/PhysRevApplied.15.024064},
	number = {2},
	urldate = {2026-09-21},
	journal = {Physical Review Applied},
	author = {Colangelo, Marco and Zhu, Di and Santavicca, Daniel F. and Butters, Brenden A. and Bienfang, Joshua C. and Berggren, Karl K.},
	month = feb,
	year = {2021},
	pages = {024064},
}

@article{zhao_superconducting_2026,
	title = {Superconducting {NbN} {Resonator} {Parametric} {Amplifiers} for {Millimetre} {Wavelengths}},
	volume = {222},
	issn = {1573-7357},
	url = {https://doi.org/10.1007/s10909-026-03432-5},
	doi = {10.1007/s10909-026-03432-5},
	number = {4},
	urldate = {2026-09-21},
	journal = {Journal of Low Temperature Physics},
	author = {Zhao, Songyuan and Withington, S. and Thomas, C. N.},
	month = jun,
	year = {2026},
	pages = {100},
}

@article{liu_temperature-dependent_2017,
	title = {Temperature-dependent performance of all-{NbN} {DC}-{SQUID} magnetometers},
	volume = {110},
	issn = {0003-6951},
	url = {https://doi.org/10.1063/1.4984300},
	doi = {10.1063/1.4984300},
	number = {22},
	urldate = {2026-09-21},
	journal = {Applied Physics Letters},
	author = {Liu, Quansheng and Wang, Huiwu and Zhang, Qiyu and Wang, Hai and Peng, Wei and Wang, Zhen},
	month = may,
	year = {2017},
	pages = {222604},
}

@misc{colletta_gap_2026,
	title = {Gap {Engineered} {Superconducting} {Multilayer} {Nanobridge} {Josephson} {Junctions}},
	url = {http://arxiv.org/abs/2603.20757},
	doi = {10.48550/arXiv.2603.20757},
	urldate = {2026-09-21},
	publisher = {arXiv},
	author = {Colletta, Giuseppe and Johny, Susan and Feng, Hua and Alkhalidi, Mohammed and Collins, Jonathan A. and Weides, Martin},
	month = mar,
	year = {2026},
	note = {arXiv:2603.20757 [cond-mat.supr-con]},
}

@article{marsili_single-photon_2011,
	title = {Single-{Photon} {Detectors} {Based} on {Ultranarrow} {Superconducting} {Nanowires}},
	volume = {11},
	issn = {1530-6984},
	url = {https://doi.org/10.1021/nl2005143},
	doi = {10.1021/nl2005143},
	number = {5},
	urldate = {2026-09-21},
	journal = {Nano Letters},
	author = {Marsili, Francesco and Najafi, Faraz and Dauler, Eric and Bellei, Francesco and Hu, Xiaolong and Csete, Maria and Molnar, Richard J. and Berggren, Karl K.},
	month = may,
	year = {2011},
	pages = {2048--2053},
}

@article{xie_brief_2023,
	title = {A brief review on the recent development of phonon engineering and manipulation at nanoscales},
	volume = {6},
	issn = {2631-7990},
	url = {https://dx.doi.org/10.1088/2631-7990/acfd68},
	doi = {10.1088/2631-7990/acfd68},
	number = {1},
	urldate = {2025-07-07},
	journal = {International Journal of Extreme Manufacturing},
	publisher = {IOP Publishing},
	author = {Xie, Siqi and Zhu, Hongxin and Zhang, Xing and Wang, Haidong},
	month = oct,
	year = {2023},
	pages = {012007},
}

@article{ma_phonon_2022,
	title = {Phonon {Engineering} of {Micro}‐ and {Nanophononic} {Crystals} and {Acoustic} {Metamaterials}: {A} {Review}},
	volume = {3},
	issn = {2688-4046},
	shorttitle = {Phonon {Engineering} of {Micro}‐ and {Nanophononic} {Crystals} and {Acoustic} {Metamaterials}},
	url = {https://www.ncbi.nlm.nih.gov/pmc/articles/PMC11936052/},
	doi = {10.1002/smsc.202200052},
	number = {1},
	urldate = {2025-09-04},
	journal = {Small Science},
	author = {Ma, Jihong},
	month = nov,
	year = {2022},
	pages = {2200052},
}

@article{yang_using_2007,
	title = {Using high-contrast salty development of hydrogen silsesquioxane for sub-10-nm half-pitch lithography},
	volume = {25},
	issn = {1071-1023, 1520-8567},
	url = {https://pubs.aip.org/jvb/article/25/6/2025/592120/Using-high-contrast-salty-development-of-hydrogen},
	doi = {10.1116/1.2801881},
	number = {6},
	urldate = {2025-08-17},
	journal = {Journal of Vacuum Science \& Technology B: Microelectronics and Nanometer Structures Processing, Measurement, and Phenomena},
	author = {Yang, Joel K. W. and Berggren, Karl K.},
	month = nov,
	year = {2007},
	pages = {2025--2029},
}

@article{vafapour_sensing_2021,
	title = {Sensing, {Switching} and {Modulating} {Applications} of a {Superconducting} {THz} {Metamaterial}},
	volume = {21},
	issn = {1558-1748},
	url = {https://ieeexplore.ieee.org/document/9402819},
	doi = {10.1109/JSEN.2021.3073078},
	number = {13},
	urldate = {2026-09-21},
	journal = {IEEE Sensors Journal},
	author = {Vafapour, Zohreh and Dutta, Mitra and Stroscio, Michael A.},
	month = jul,
	year = {2021},
	pages = {15187--15195},
}

@article{mirhosseini_superconducting_2018,
	title = {Superconducting metamaterials for waveguide quantum electrodynamics},
	volume = {9},
	copyright = {2018 The Author(s)},
	issn = {2041-1723},
	url = {https://www.nature.com/articles/s41467-018-06142-z},
	doi = {10.1038/s41467-018-06142-z},
	number = {1},
	urldate = {2026-09-21},
	journal = {Nature Communications},
	publisher = {Nature Publishing Group},
	author = {Mirhosseini, Mohammad and Kim, Eunjong and Ferreira, Vinicius S. and Kalaee, Mahmoud and Sipahigil, Alp and Keller, Andrew J. and Painter, Oskar},
	month = sep,
	year = {2018},
	pages = {3706},
}

@article{xu_achieving_2010,
	title = {Achieving the {Theoretical} {Depairing} {Current} {Limit} in {Superconducting} {Nanomesh} {Films}},
	volume = {10},
	issn = {1530-6984},
	url = {https://doi.org/10.1021/nl102584j},
	doi = {10.1021/nl102584j},
	number = {10},
	urldate = {2026-09-21},
	journal = {Nano Letters},
	author = {Xu, Ke and Cao, Peigen and Heath, James R.},
	month = oct,
	year = {2010},
	pages = {4206--4210},
}

@article{sreenivasan_nanoimprint_2017,
	title = {Nanoimprint lithography steppers for volume fabrication of leading-edge semiconductor integrated circuits},
	volume = {3},
	copyright = {2017 The Author(s)},
	issn = {2055-7434},
	url = {https://www.nature.com/articles/micronano201775},
	doi = {10.1038/micronano.2017.75},
	number = {1},
	urldate = {2026-09-21},
	journal = {Microsystems \& Nanoengineering},
	publisher = {Nature Publishing Group},
	author = {Sreenivasan, S. V.},
	month = sep,
	year = {2017},
	pages = {17075},
}

@article{zhao_fabrication_2015,
	title = {Fabrication of {Nb} {Superconducting} {Nanowires} by {Nanoimprint} {Lithography}},
	volume = {25},
	issn = {1558-2515},
	url = {https://ieeexplore.ieee.org/document/6990535},
	doi = {10.1109/TASC.2014.2382976},
	number = {3},
	urldate = {2026-09-21},
	journal = {IEEE Transactions on Applied Superconductivity},
	author = {Zhao, Lu and Jin, Yirong and Li, Jie and Deng, Hui and Li, Hekang and Huang, Keqiang and Cui, Limin and Zheng, Dongning},
	month = jun,
	year = {2015},
	pages = {1--5},
}

@article{jeong_directed_2013,
	title = {Directed self-assembly of block copolymers for next generation nanolithography},
	volume = {16},
	issn = {1369-7021},
	url = {https://www.sciencedirect.com/science/article/pii/S1369702113003866},
	doi = {10.1016/j.mattod.2013.11.002},
	number = {12},
	urldate = {2026-09-21},
	journal = {Materials Today},
	author = {Jeong, Seong-Jun and Kim, Ju Young and Kim, Bong Hoon and Moon, Hyoung-Seok and Kim, Sang Ouk},
	month = dec,
	year = {2013},
	pages = {468--476},
}

@article{robbins_block_2016,
	title = {Block copolymer self-assembly–directed synthesis of mesoporous gyroidal superconductors},
	volume = {2},
	url = {https://www.science.org/doi/10.1126/sciadv.1501119},
	doi = {10.1126/sciadv.1501119},
	number = {1},
	urldate = {2026-09-21},
	journal = {Science Advances},
	publisher = {American Association for the Advancement of Science},
	author = {Robbins, Spencer W. and Beaucage, Peter A. and Sai, Hiroaki and Tan, Kwan Wee and Werner, Jörg G. and Sethna, James P. and DiSalvo, Francis J. and Gruner, Sol M. and Van Dover, Robert B. and Wiesner, Ulrich},
	month = jan,
	year = {2016},
	pages = {e1501119},
}

@article{tatarintsev_effect_2020,
	title = {The {Effect} of {Temperature} on the {Development} of a {Contrast} {HSQ} {Electronic} {Resist}},
	volume = {49},
	issn = {1608-3415},
	url = {https://doi.org/10.1134/S1063739720030063},
	doi = {10.1134/S1063739720030063},
	number = {3},
	urldate = {2026-09-21},
	journal = {Russian Microelectronics},
	author = {Tatarintsev, A. A. and Shishlyannikov, A. V. and Rudenko, K. V. and Rogozhin, A. E. and Ieshkin, A. E.},
	month = may,
	year = {2020},
	pages = {151--156},
}

@article{gorelick_direct_2010,
	title = {Direct e-beam writing of dense and high aspect ratio nanostructures in thick layers of {PMMA} for electroplating},
	volume = {21},
	issn = {0957-4484},
	url = {https://doi.org/10.1088/0957-4484/21/29/295303},
	doi = {10.1088/0957-4484/21/29/295303},
	number = {29},
	urldate = {2026-09-21},
	journal = {Nanotechnology},
	author = {Gorelick, Sergey and Guzenko, Vitaliy A and Vila-Comamala, Joan and David, Christian},
	month = jul,
	year = {2010},
	pages = {295303},
}

@article{bose_review_2023,
	title = {A review of superconductivity in nanostructures—from nanogranular films to anti-dot arrays},
	volume = {36},
	issn = {0953-2048},
	url = {https://doi.org/10.1088/1361-6668/acc980},
	doi = {10.1088/1361-6668/acc980},
	number = {6},
	urldate = {2026-09-23},
	journal = {Superconductor Science and Technology},
	publisher = {IOP Publishing},
	author = {Bose, Sangita},
	month = may,
	year = {2023},
	pages = {063003},
}

@article{kroll_magnetic-field-resilient_2019,
	title = {Magnetic-{Field}-{Resilient} {Superconducting} {Coplanar}-{Waveguide} {Resonators} for {Hybrid} {Circuit} {Quantum} {Electrodynamics} {Experiments}},
	volume = {11},
	url = {https://link.aps.org/doi/10.1103/PhysRevApplied.11.064053},
	doi = {10.1103/PhysRevApplied.11.064053},
	number = {6},
	urldate = {2026-09-23},
	journal = {Physical Review Applied},
	publisher = {American Physical Society},
	author = {Kroll, J.G. and Borsoi, F. and van der Enden, K.L. and Uilhoorn, W. and de Jong, D. and Quintero-Pérez, M. and van Woerkom, D.J. and Bruno, A. and Plissard, S.R. and Car, D. and Bakkers, E.P.A.M. and Cassidy, M.C. and Kouwenhoven, L.P.},
	month = jun,
	year = {2019},
	pages = {064053},
}

@article{tettamanzi_superconducting_2009,
	title = {Superconducting transition in {Nb} nanowires fabricated using focused ion beam},
	volume = {20},
	issn = {0957-4484},
	url = {https://doi.org/10.1088/0957-4484/20/46/465302},
	doi = {10.1088/0957-4484/20/46/465302},
	number = {46},
	urldate = {2026-09-23},
	journal = {Nanotechnology},
	author = {Tettamanzi, G C and Pakes, C I and Potenza, A and Rubanov, S and Marrows, C H and Prawer, S},
	month = oct,
	year = {2009},
	pages = {465302},
}

@article{nam_contrast_2009,
	title = {Contrast enhancement behavior of hydrogen silsesquioxane in a salty developer},
	volume = {27},
	issn = {1071-1023},
	url = {https://doi.org/10.1116/1.3245991},
	doi = {10.1116/1.3245991},
	number = {6},
	urldate = {2026-09-23},
	journal = {Journal of Vacuum Science \& Technology B: Microelectronics and Nanometer Structures Processing, Measurement, and Phenomena},
	author = {Nam, Sung-Wook and Rooks, Michael J. and Yang, Joel K. W. and Berggren, Karl K. and Kim, Hyun-Mi and Lee, Min-Hyun and Kim, Ki-Bum and Sim, Jae Hwan and Yoon, Do Yeung},
	month = nov,
	year = {2009},
	pages = {2635--2639},
}

@article{mohammadi-arzanagh_enhanced_2025,
	title = {Enhanced {Cooper} pairing in nanopatterned metals},
	volume = {112},
	url = {https://link.aps.org/doi/10.1103/v82c-tmdc},
	doi = {10.1103/v82c-tmdc},
	number = {10},
	urldate = {2026-09-23},
	journal = {Physical Review B},
	publisher = {American Physical Society},
	author = {Mohammadi-Arzanagh, Masoud and Grankin, Andrey and Galitski, Victor and Hafezi, Mohammad},
	month = sep,
	year = {2025},
	pages = {L100509},
}

@article{zhang_improving_2014,
	title = {Improving the adhesion of hydrogen silsesquioxane ({HSQ}) onto various substrates for electron-beam lithography by surface chemical modification},
	volume = {128},
	issn = {0167-9317},
	url = {https://www.sciencedirect.com/science/article/pii/S0167931714002202},
	doi = {10.1016/j.mee.2014.05.027},
	urldate = {2026-09-24},
	journal = {Microelectronic Engineering},
	author = {Zhang, Zhiqiang and Duan, Huigao and Wu, Yihui and Zhou, Wuping and Liu, Cong and Tang, Yuguo and Li, Haiwen},
	month = oct,
	year = {2014},
	pages = {59--65},
}

@article{radtke_reliable_2019,
	title = {Reliable {Nanofabrication} of {Single}-{Crystal} {Diamond} {Photonic} {Nanostructures} for {Nanoscale} {Sensing}},
	volume = {10},
	issn = {2072-666X},
	url = {https://pmc.ncbi.nlm.nih.gov/articles/PMC6915366/},
	doi = {10.3390/mi10110718},
	number = {11},
	urldate = {2026-09-24},
	journal = {Micromachines},
	author = {Radtke, Mariusz and Nelz, Richard and Slablab, Abdallah and Neu, Elke},
	month = oct,
	year = {2019},
	pages = {718},
}

@article{chen_effects_2006,
	series = {Micro- and {Nano}-{Engineering} {MNE} 2005},
	title = {Effects of developing conditions on the contrast and sensitivity of hydrogen silsesquioxane},
	volume = {83},
	issn = {0167-9317},
	url = {https://www.sciencedirect.com/science/article/pii/S0167931706002607},
	doi = {10.1016/j.mee.2006.01.167},
	number = {4},
	urldate = {2026-09-24},
	journal = {Microelectronic Engineering},
	author = {Chen, Yifang and Yang, Haifang and Cui, Zheng},
	month = apr,
	year = {2006},
	pages = {1119--1123},
}

@article{ocola_nanoscale_2009,
	title = {Nanoscale geometry assisted proximity effect correction for electron beam direct write nanolithography},
	volume = {27},
	issn = {1071-1023},
	url = {https://doi.org/10.1116/1.3237135},
	doi = {10.1116/1.3237135},
	number = {6},
	urldate = {2026-09-24},
	journal = {Journal of Vacuum Science \& Technology B: Microelectronics and Nanometer Structures Processing, Measurement, and Phenomena},
	author = {Ocola, L. E.},
	month = nov,
	year = {2009},
	pages = {2569--2571},
}

@article{glover_conductivity_1957,
	title = {Conductivity of {Superconducting} {Films} for {Photon} {Energies} between 0.3 and 40 {kTc}},
	volume = {108},
	url = {https://link.aps.org/doi/10.1103/PhysRev.108.243},
	doi = {10.1103/PhysRev.108.243},
	number = {2},
	urldate = {2026-10-26},
	journal = {Physical Review},
	publisher = {American Physical Society},
	author = {Glover, R. E. and Tinkham, M.},
	month = oct,
	year = {1957},
	pages = {243--256},
}

@article{gangnaik_new_2017,
	title = {New {Generation} {Electron} {Beam} {Resists}: {A} {Review}},
	volume = {29},
	issn = {0897-4756},
	shorttitle = {New {Generation} {Electron} {Beam} {Resists}},
	url = {https://doi.org/10.1021/acs.chemmater.6b03483},
	doi = {10.1021/acs.chemmater.6b03483},
	number = {5},
	urldate = {2026-11-01},
	journal = {Chemistry of Materials},
	author = {Gangnaik, Anushka S. and Georgiev, Yordan M. and Holmes, Justin D.},
	month = mar,
	year = {2017},
	pages = {1898--1917},
}

@article{nuss_dynamic_1991,
	title = {Dynamic conductivity and coherence peak in {YBa} 2 {Cu} 3 {O} 7 superconductors},
	volume = {66},
	copyright = {http://link.aps.org/licenses/aps-default-license},
	issn = {0031-9007},
	url = {https://link.aps.org/doi/10.1103/PhysRevLett.66.3305},
	doi = {10.1103/PhysRevLett.66.3305},
	number = {25},
	urldate = {2026-11-02},
	journal = {Physical Review Letters},
	author = {Nuss, Martin C. and Mankiewich, P. M. and O’Malley, M. L. and Westerwick, E. H. and Littlewood, Peter B.},
	month = jun,
	year = {1991},
	pages = {3305--3308},
}

\end{document}